\documentclass[aps,prd,twocolumn,a4paper]{revtex4-1}
\usepackage{ulem}
\usepackage{color}
\usepackage{graphicx}
\usepackage{epsfig}
\usepackage{mathrsfs}
\newcommand{\be}{\begin{equation}}
\newcommand{\ee}{\end{equation}}
\newcommand{\ba}{\begin{eqnarray}}
\newcommand{\ea}{\end{eqnarray}}
\newcommand{\nn}{\nonumber\\}
\begin{document}
\title{ Impact of momentum anisotropy and turbulent chromo-fields on thermal particle production in quark-gluon plasma medium}
\author{Vinod Chandra$^a,$}
\email{vchandra@iitgn.ac.in}
\author{V. Sreekanth$^b$}
\email{sreekanth@chep.iisc.ac.in}
\affiliation{$^{a}$ Indian Institute of Technology Gandhinagar, Gandhinagar-382355, Gujarat, India}
\affiliation{$^b$ Centre for High Energy Physics, Indian Institute of Science, Bangalore 560012, India}
\date{\today}
\begin{abstract}
Momentum anisotropy present during the hydrodynamic evolution of 
Quark-Gluon Plasma (QGP) in RHIC may lead to chromo-Weibel instability and 
turbulent chromo-fields.The dynamics of the quark and gluon momentum distributions in this case 
is governed by an effective  diffusive  Vlasov equation (linearized).
The solution of this linearized transport equation 
for the modified momentum distribution functions
lead to the mathematical form of non-equilibrium momentum distribution
functions of quarks/antiquarks and gluons. The modification to 
these distributions encode the physics of turbulent color fields and momentum anisotropy.
In the present manuscript, we employ these  distribution functions to 
to estimate  thermal dilepton production rate in the QGP medium. The production rate is seen to have 
appreciable sensitivity to the strength of  the anisotropy.

\vspace{2mm}

\vspace{2mm}
\end{abstract}
\maketitle

\section{Introduction}
The experimental observation from the relativistic heavy-ion collisions  at RHIC, BNL and LHC CERN, have
strongly suggested the creation of quark-gluon-plasma (QGP) in a near perfect fluid state~\cite{expt,expt1}.
The space time dynamics of the QGP has been modelled within the framework of  
second order relativistic dissipative hydrodynamics~\cite{echo,sch,bozek,review1,review2,song,niemi}.
The hydrodynamical predictions for the collective flow coefficients and particle spectra in heavy-ion collisions (HIC), 
seen to work well for hadronic probes. The role of hydrodynamics in HIC has been to convert the geometrical fluctuations 
in the initial geometry of the reaction zone (soon after the collisions) to the 
momentum anisotropy, which finally leads to collective flow in the hadronic observables. Therefore,
the momentum anisotropy has been there during the entire space-time evolution of the QGP. 
On the other hand, the early stages of the  HIC  where the momentum distribution is far from equilibrium and 
highly anisotropic, leading to instabilities has been well explored by several authors~\cite{instab2-4} along with the detailed study on its consequences. For a very recent review, 
we refer readers to the Ref.~\cite{strik_pr}.

 Based on the experimental observations, the QGP turned out to be a near perfect fluid with  a tiny value for its shear viscosity to entropy ratio ($\eta/S\sim 1/{4\pi}$, which is smallest among almost all the known fluids in nature).
It has been realized that collisional processes among effective gluonic and quark degrees of freedom of the QGP alone could not explain such a small number. The presence of momentum anisotropy is seen to play dominant role in 
 in  substantial modulation of shear viscosity of the hot QCD medium in weak coupling regime~\cite{bmuller, chandra_eta} and may provide a possible explanation for the small, $\eta/S$.
The main focus here is on the momentum anisotropy  in the later stages of the 
collisions where the dynamics is governed by the hydrodynamics while the matter is near equilibrated.
The physics of such momentum anisotropy is quite crucial to understand the QGP medium as in certain cases, 
it may lead to chromo-Weibel instability in the QGP~\cite{chromoweibel} and turbulent chromo-fields~\cite{bmuller}. 
As mentioned in~\cite{peter}, the fields generated by such instabilities help in 
rapid isotropization of the parton distributions and drive the system to hydrodynamical regime.

Instabilities have been extensively studied in the context of classical Yang-Mills dynamics and their role in thermalization and isotropization of the system~\cite{strik_pr} and their role in setting up the turbulence in the plasma.  The two main frameworks for
these investigations have been the classical-statistical lattice~\cite{bergers12,bergers14} and CGC frameworks~\cite{fukushima,gelis}. 
These classical-statistical lattice simulations~\cite{bergers12}  revealed that after an initial   transient regime dominated by plasma instabilities and free streaming, the non-Abelian plasma exhibits the universal self-similar dynamics characteristic of wave turbulence.
To handle the static case,  Kurkela and Moore ~\cite{kuk2012a} proposed a new algorithm for solving the Yang-Mills 
equations in an expanding box that takes care of the transverse dynamics without getting affected by the longitudinal coarse lattices. The turbulence phenomenon 
has been studied in static box by Kurkela and Moore~\cite{kukmo} and Berges et al.~\cite{bergers12} and they observed a cascade of energy flow towards higher momenta  and existence  of a scaling solution therein.
On the other hand, the CGC lattice simulations  by  Fukushima~\cite{fukushima}  and Fukushima and Gelis~\cite{gelis}  have seen  an energy flow from low to high wave number modes which eventually 
resulting in a spectrum consistent with Kolmogorov's power law form indication of turbulent behavior. 
The later stages of longitudinally expanding plasmas after instabilities have stopped growing was studied by Berges et al.~\cite{bergers14} where they highlighted the role of quantum fluctuations in the classical lattice simulations and their role 
in deciding the time scales for the system to isotropize and approache thermal equilibrium. From all these above studies and a few more (for details we refer the reader to~\cite{strik_pr}), it is to be noted that the instabilities may lead to plasma turbulence.
The role of instabilities and plasma turbulence might play crucial role in understanding the properties of the QGP in heavy-ion collisions while one concentrates on the anisotropy (momentum) in the later stages of the collisions.

The physics of the chromo-Weibel instability
(non-Abelian analogue of Weibel instability~\cite{weibel}) might play
crucial role in understanding the space-time evolution and properties of QGP medium.
The momentum anisotropy present during the hydrodynamic expansion of the QGP induces
instability to the Yang-Mills field  equations. The Weibel type of  instabilities 
can be seen in the  expanding QGP, since the  width of the momentum component in the direction
of the expansion squeezes by the expansion, leading to an anisotropic momentum distribution. 
The instability in the rapidly expanding QGP in heavy ion collisions may also lead to the 
plasma turbulence~\cite{bmuller}. Note that the plasma turbulence describes a random, non-thermal pattern of
excitation of coherent color field modes in the QGP. The power spectrum, in this case, turns out to be
similar to that of vortices in a turbulent fluid~\cite{bmuller}.

An effective transport equation (Vlasov-Boltzmann) has been setup in~\cite{bmuller} in a form 
applicable to the case of the turbulent QGP by making additional assumptions regarding the 
field distributions in the Vlasov term. To reflect the turbulent nature certain spatio-temporal correlation 
structure for fields at different 
space-time points (Gaussian form in~\cite{bmuller}) has been considered. This allows one to rewrite the 
color-octet particle distribution function 
in the form of dissipative term acting on the singlet distribution and 
eventually leads to diffusive Vlasov-Boltzmann equation. 
Notably, the Vlasov (Force term) operator, thus, obtained in the effective transport equation 
only picks up the contributions from the anisotropy. 
The color electric field contribute through the thermal conductivity. The color-magnetic field mainly picks
the anisotropy in the plasma medium. Here, we are only dealing with the latter. It has already been realized that such 
turbulent color fields,  may contribute significantly to the transport processes in the QGP.  
As there is a significant decrease in the transport coefficients in the presence of 
turbulent fields~\cite{dupree,bmuller}, the small shear viscosity to entropy ratio ($\eta/S$)
can perhaps be understood in terms of the dominance of such fields. 

The prime goal here is to investigate the dilepton
production rate in the presence of 
chromo-Weibel instability. This could be done by first modeling  the non-equilibrium momentum distribution 
functions~\cite{vc_hq} that describe expanding anisotropic QGP followed by employing it to the kinetic theory description
of dilepton production in the QGP medium. Our formalism  is a straightforward extension  of the Ref.~\cite{bmuller} for the 
interacting/realistic hot QCD equation of state (here the (2+1)-flavor lattice EOS described in terms of a quasi-particle model). The hot QCD 
medium effects enter through the quasi-particle distribution functions along with the non-trivial energy dispersions. The near equilibrium  quasi-quark and quasi-gluon distributions 
that are employed here are obtained earlier in~\cite{vc_hq} and utilized in the context of studying heavy-quark dynamics in the anisotropic QGP/QCD medium.
In this work, we employ them in exploring the dilepton production from the thermal QGP medium following the kinetic theory description of the dilepton production 
by $q\bar{q}$ annihilation at leading order within (1+1)-d boost invariant expansion of the thermal QGP medium in longitudinal direction.

This is perhaps the first time,  the impact of momentum anisotropy induced turbulent color fields has been 
included in the thermal particle  production  in hot QCD/QGP medium. As it will be seen in the later part 
of the manuscript that such effects indeed play significant role and can not simply be ignored for the momenta and temperatures 
accessible in RHIC. It is to be mentioned that such anisotropy induced momentum distributions have 
already been exploited to investigate the heavy-quark dynamics in the QGP medium ~\cite{vc_hq} . 
The heavy quark dynamics gets significant impact from  such effects. Such effects may also be helpful 
to resolve the simultaneous estimation of $v_2$ and $R_{AA}$ for the heavy-quarks~\cite{sdas}.

Particles are produced from the thermal medium of expanding fireball created in heavy-ion collisions, 
throughout its evolution, carrying crucial information about 
the state of the constituent matter \cite{Alam:1996fd,Alam:1999sc,Peitzmann:2001mz,vanHees:2007th}.
Effects of equation of states and non-equilibrium scenarios like dissipation {\it etc.} on such thermal 
particles from the QGP phase have been studied by various authors 
\cite{McLerran:1984ay,Bhatt:2009zg,Bhatt:2010cy,PeraltaRamos:2010cw,Bhatt:2011kx,Bhalerao:2013aha,Chandra:2015rdz}. 
In this context it will be interesting to look how the presence of momentum anisotropy and  turbulent
chromo-fields will be affecting the thermal particle production.

The paper is organized as follows.  In Section. II, the modeling   of non-equilibrium distribution function
in the presence of anisotropy driven instability has been 
presented. Section III,  deals with the thermal particle 
productions rates in the anisotropic background medium and
Section IV, discusses the yields in the presence of expanding 
medium in heavy-ion collisions. In section V, conclusions 
and outlook have been presented.

\section{Near (Non)-equilibrium quark and gluon distributions}
 Recall, the momentum anisotropy present in quark and gluon momentum distribution functions 
 induces instability in the Yang-Mills equations in similar way as Weibel instability in the case of 
 Electromagnetic  plasmas. This instability while coupled with the rapid expansion of the QGP leads to anomalous transport and 
 modulates the transport coefficients of the plasma substantially. This fact is realized by Dupree in the case of 
Electro-Magnetic plasmas in 1954~\cite{dupree} and
 later generalized for the non-Abelian plasmas in Refs.~\cite{bmuller,chandra_eta1}. In the context of QGP, the phenomenon of the anomalous transport 
 is realized at the later stages of the collisions, as due to the hydrodynamic expansion of the QGP, one  has appreciable momentum anisotropy present in 
 thermal distribution functions of quark and gluons. 
 
 To obtain the near equilibrium distribution within linearized transport theory we first need to have an adequate 
 model for the isotropic (equilibrium) momentum distributions functions for the QGP degrees of freedom. To that end, we employ 
 a quasi-particle description~\cite{chandra_quasi} of the lattice based  QGP equation of state ~\cite{cheng,leos1_lat}. In this model,
 form of the equilibrium distribution functions,  $ f_{eq}$ are obtained by encoding the strong interaction effects 
 in terms of effective fugacities for quarks/gluons ($z_{g,q}$) as:
\ba
\label{eq1}
f_0^{g/q} &=& \frac{z_{g/q}\exp[-\beta E_p]}{\bigg(1\mp z_{g/q}\exp[-\beta E_p]\bigg)},\nn
\ea
where $p=|\vec{p}|$,  $E_p=p$ for gluons and light quarks ($u$ and $d$).  On the other hand for strange quark in (2+1)-flavor QCD, $E_p=\sqrt{p^2+m_s^2}$ for (s-quarks). 
Here, $m_s$ denotes the mass of the strange quark, and $\beta=T^{-1}$ denotes inverse of the 
temperature. Since, the model is valid for temperatures that are higher than  $T_c$, hence, we ignore the strange quark mass effects.  The sub/superscript q denotes the $u$, $d$ and $s$ quarks.
The effective fugacities ($z_{g/q}$)  the model are not merely a temperature
dependent parameters that encode the hot QCD medium effects;  they lead to 
non-trivial dispersion relation both in the gluonic and quark sectors as,
\ba
\label{eq2}
\omega_{g,q}&=&E_p+T^2\partial_T ln(z_{g,q}).
\ea
For more detailed discussion on the understanding of $z_{g,q}$, we refer the reader to Ref.~\cite{chandra_quasi}.  This quasi-particle description of hot QCD medium 
has seen to be highly useful in understanding the transport properties of hot QCD medium~\cite{chandra_eta, chandra_eta1,chandra_zeta}, dilepton production in the QGP medium~\cite{Chandra:2015rdz},  electrical conductivity and
 charge diffusion in hot QCD medium~\cite{chandra_sukanya}.

It is worth to mention that  there have been  other quasi-particle descriptions in the literature, those could be 
characterized as, effective mass models ~\cite{effmass1,effmass2},
effective mass models with gluon condensate~\cite{effmass_glu}, and 
effective models with Polyakov loop~\cite{effective_pol,effmass_pol}.  The effective model with Polyakov loop 
in~\cite{effmass_pol} has been thermodynamically consistent.  The  model  employed here, is fundamentally distinct from all these models.
In the presence of non-trivial quasi-particle  dispersions (as in case of the say effective mass model or our model), the kinetic theory definition of 
energy-momentum tensor, $T^{\mu\nu}$ will get modified in terms of capturing the medium dependent terms~\cite{jeon_yaffe, blumn_1, dusling, chandra_quasi}. 
Such modifications are mandated by the fact that  the $T^{\mu\nu}$  must incorporate the effects of trace anomaly. This fact in the case of effective mass models 
have been described and an effective $T^{\mu\nu}$ is obtained in  Ref.~\cite{blumn_1}. The authors further, estimated the viscosities of hot QCD matter~\cite{blumn_1}.
There has been more detailed study in this direction~\cite{kapusta_pn}.
The  mathematical expression for  the modified  $T^{\mu\nu}$ has been  
obtained  for the current model in Ref.~\cite{chandra_zeta}. It is further to be noted that, Bluhm {\it et al.}~\cite{blumn_2}, highlighted the utility of the effective mass models for the 
relativistic heavy ion collisions  where non-zero baryon density aspects are also explored.

Next, we set-up an effective transport equation for the near-equilibrium momentum distribution functions for quarks and gluons in the 
presence of initial momentum anisotropy and space time expansion of the QGP.

 \subsection{An Effective kinetic equation--the Dupree-Vlasov equation}
We start with the following ansatz for the non-equilibrium distribution function
 \begin{equation}
 \label{eqf}
  f(\vec{p}, \vec{r})=\frac{z_{g,q} \exp(-\beta u^\mu p_\mu)}{1\pm z_{g,q} \exp\big(-\beta u^\mu p_\mu +f_1 (\vec{p}, \vec{r})\big)},
 \end{equation}
 where $z_{g,q}$ are the effective gluon, quark fugacities coming from the isotropic modeling of the QGP in 
 terms of lattice QCD equation of state 
 and $u^\mu$ is the fluid 4-velocity considering fluid picture of the QGP medium. 
 Here, $f_1 (\vec{p}, \vec{r})$ denotes the effects from the anisotropy (momentum) 
to the equilibrium distribution function. To achieve the above mentioned near equilibrium situation, 
 $f_1$ must be a small perturbation. Under this condition, we obtain,
 \begin{equation}
  f(\vec{p}, \vec{r})=f_0(p)+f_0 (1\pm f_0 (p)) f_1(\vec{p},\vec{r}) +O(f_1(\vec{p}, \vec{r})^2).
 \end{equation}
The {\it plus} sign is for gluons and {\it minus} sign is for the quarks/antiquarks.

Next, the following form for the ansatz is considered for the linear order  perturbation to the isotropic  gluon and quarks distribution functions respectively,
\begin{equation}
f_1 (\vec{p},\vec{r})= -\frac{1}{\omega_{g,q} T^2}\bigg(  p_i p_j  \Delta(\vec{p}) (\nabla u)_{ij}
\bigg),
\end{equation}
where $f_1 (\vec{p},\vec{r})\equiv f_1^{g,q}$ and $\Delta \equiv \Delta^{g,q}$.
The quantity $\Delta$ captures effects from the momentum anisotropy.
In the local rest frame of the fluid (LRF) $f_0=f_{eq}=(f_0^g, f_0^q)$, and considering longitudinal boost invariance
\cite{Bjorken:1982qr}, 
we obtain,
$\nabla\cdot \vec{u}=\frac{1}{\tau}$ and ${\nabla u}_{ij}=\frac{1}{3 \tau} diag (-1,-1,2)$, leading to \\
\begin{equation}
 f_1^{g,q}=-\frac{\Delta^{g,q} (p)}{\omega_{g,q} T^2 \tau} (p_z^2-\frac{p^2}{3}).
\end{equation}
Let us now proceed to set up the effective transport equation in the presence of turbulent chromo-fields that are induced by the 
momentum anisotropy in the thermal distribution of the quasi-gluons and quarks while coupled with the rapid expansion of the QGP medium. 

\subsubsection{Effective transport equation in turbulent chromo fields}
 The near-equilibrium (anisotropic) momentum distributions for the quasi-gluons and quasi-quarks  in our case 
were obtained by solving the Vlasov-Boltzmann equation in the presence of turbulent 
chromo fields in~\cite{chandra_eta}.  The approach has been   based on a straightforward extension of the 
work Asakawa {\it et. al, }~\cite{bmuller}. Below, we offer the essential steps in the determination of the distributions.

The evolution of the quasi-quark and quasi-gluon momentum distribution functions 
in the anisotropic QGP medium with color fields can be  obtained by setting up  
Vlasov-Boltzmann equation~\cite{Heinz:1983nx} as:
\begin{equation}
v^\mu\frac{\partial}{\partial x^\mu} f({\bf r},{\bf p},t) 
+ g {\bf F}^a\cdot\nabla_q f^a({\bf r},{\bf p},t) 
= 0 \, 
\label{eqVl}
\end{equation}
where  $f({\bf r},{\bf q},t)$ represent the parton distribution in 
phase space  (sums over all parton colors),  the quantities ${\bf p}\equiv \vec{p}$ and ${\bf r}\equiv \vec{r}$. Here,   $f^a({\bf r},{\bf q},t)$ 
denotes the color octet distribution function. Note that, here we are dealing with the collisionless plasma.
Both the distributions, $f$ and $f^a$
are defined in the semi-classical formalism in Ref.~\cite{Heinz:1984yq} as the moments of the distribution 
function ${\tilde f}({\bf r},{\bf p},Q,t)$ in an extended phase space 
that includes the color sector as:
\begin{eqnarray}
f({\mathbf r},{\bf p},t) 
&=& \int dQ\, {\tilde f}({\bf r},{\bf p},Q,t) \, ,  
\\
f^a({\mathbf r},{\bf p},t) 
&=& \int dQ\, Q^a {\tilde f}({\bf r},{\bf p},Q,t) \, .
\end{eqnarray}
Here $Q^a$ denotes the color charge, $v^\mu=\frac{p^\mu}{p^0}$, $p^\mu=(p^0=E_p, \vec{p})$.  The color Lorentz force is defined as:
\begin{equation}
 {\bf F}^a=E^a+{\bf v}\times B^a.
\end{equation}
The color octet distribution function, $f^a$  will satisfy a transport equation which  involve coupling with the phase 
space distributions of higher color-SU(3) representations.  The  near equilibrium considerations allows us to 
truncate this hierarchy by keeping only  the lowest order term in the gradients for both $f$ and $f^a$.
The color octet distribution identically  vanishes at equilibrium. This  implies that it is at least linear in perturbation.  
With these consideration, the  transport equation    
for $f^a$  is obtained as~ \cite{Heinz:1983nx,Heinz:1984yq}:
\begin{equation}
v^\mu \frac{\partial f^a}{\partial x^\mu} + g f_{abc} A^b_\mu v^\mu f^c
+ \frac{g C_2}{N_c^2-1} {\bf F}^a\cdot\nabla_p f  = 0,
\label{eqVl2}
\end{equation}
where $C_2$ and $f_{abc}$ represent quadratic Casimir invariant ($C_2\equiv \big(N_c, (N_c^2-1)/{2 N_c}\big)$ 
and structure constants of the $SU(N_c)$ respectively, and $A^{\mu}$ represents the gauge field. 

Now the goal is to solve Eq.(\ref{eqVl2}) and
obtain $f^a$ in terms of $f$ and finally solve Eq.(\ref{eqVl}), in the case of turbulent Chromo fields. This has been 
done in~\cite{bmuller} treating isotropic hot QCD/QGP as the ultra-relativistic gas of the quarks-antiquarks and gluons. In our case, the isotropic (local equilibrium) state is described in terms of 
effective quasi-gluon and quasi-quarks/antiquarks that describes the realistic hot QCD EOS (lattice) in the local rest frame of the QGP fluid. The extension of the whole treatment to 
the present case  turned out to be quite straightforward. The interactions  enter through the distribution functions and modified energy dispersions~\cite{chandra_eta}. 
Following~\cite{bmuller, chandra_eta}, we obtain an effective  diffusive Vlasov-Boltzmann 
equation for the turbulent fields:
\begin{equation}
v^\mu\frac{\partial}{\partial x^\mu} \bar{f} 
-{\mathcal F_V} \bar{f}= 0 \, .
\label{eq:DVBE}
\end{equation}
where,
\begin{eqnarray}
 {\mathcal F_V}
&=& - \frac{g^2 C_2}{4(N_c^2-1)\omega_{g,q}^2}\langle E^2 +B^2 \rangle\, 
    \tau_{\rm m}\nonumber\\&&\times
    \left[({\mathbf L}^{(p)})^2 - (L^{(p)}_z)^2\right] \, 
\label{eq:D-1b}
\end{eqnarray}
Here, $-i{\mathbf p}\times\nabla_p = {\mathbf L}^{(p)}$,
${\bf v}={\bf p}/\omega_{g,q}$ and for the  $\bar{f}(\vec{p})$ we shall employ 
$f(\vec{p})$ in Eq.(\ref{eq1}).
Here, $\bar{f}$ denotes the ensemble averaged momentum distribution (singlet) function of quasi-partons~\cite{bmuller}. In our case, 
$\bar{f}\equiv f(\vec{p}, \vec{r})$, as given in Eq. (\ref{eqf}). Note that we are only considering the anomalous transport 
and the collision term is not taken in to account here.  The argument here is based on the fact that the anomalous transport process, which leads to 
highly significant suppression of the transport coefficients in the expanding QGP, is the dominant mechanism to understand the tiny 
value of the $\eta/S$ for the QGP. We intend to  revisit  it, employing an appropriate collision term in the near future.
In  obtaining the above mentioned diffusive Vlasov equation, appropriate forms correlation functions functions (Gaussian correlators  satisfy all the assumptions) for the color fields have been chosen~\cite{bmuller} while 
assuming that the color-electric and -magnetic field are uncorrelated. Being real and symmetry properties of the Gaussian correlators with respect to two space-time points express the chaotic nature of the hot QCD plasma. This is how the turbulent nature of the plasma and its 
effects on the dynamics through the transport equation is introduced here, following Ref.\cite{bmuller}.

The force term (${\mathcal F_V}$) in the case of chromo-electromagnetic plasma will have the following form~\cite{bmuller}:
\begin{eqnarray}
 {\mathcal F_V} \bar{f}(p) &&\equiv {\mathcal F_V} f(\vec{p}, \vec{r})\nonumber\\
 &=&   \frac{g^2 C_2}{3 (N_c^2-1) \omega^2_{g,q}} \langle E^2+B^2 \rangle \tau_m\nonumber\\ &&\times [({\mathbf L}^{(p)})^2 - (L^{(p)}_z)^2]
f_{eq}(1\pm f_{eq}) p_i p_j (\nabla u)_{ij}.
 \end{eqnarray}
 The quantities $\langle E^2\rangle$ and $\langle B^2 \rangle$
are the color averaged chromo-electric and chromo-magnetic fields (average over the ensemble of turbulent color fields~\cite{bmuller}), 
$\tau_m$ is the time scale  for the instability. 

Now, the action of the drift operator on $f_{eq}$ is given by:
\begin{eqnarray}
\label{eqd}
(v\cdot\partial)f_{eq}&=&-f_{eq}(1+f_{eq})\bigg\lbrace(p-\partial_{\beta} \ln(z_{g,q}))v\cdot \partial (\beta)
\nonumber\\&&+\beta (v\cdot \partial)(u\cdot p)\bigg\rbrace,
\end{eqnarray}
 where,  $p-\partial_{\beta} \ln(z_{g/q})\equiv \omega_{g,q}$,  is the modified dispersion relations.
After some mathematical massaging, we obtain the following expression for the drift term, 
\begin{eqnarray}
\label{eqd} 
(v\cdot\partial)f_{eq}(p)&=&f_{eq}(1\pm f_{eq})\bigg[\frac{p_i p_j}{\omega_{g,q} T} (\nabla u)_{ij}\nonumber\\&&-\frac{m^2_D \langle E^2\rangle \tau^{el}
\omega_{g,q}}{3T^2 {\partial {\varepsilon}}/{\partial T}}\nonumber\\&&+(\frac{p^2}{3\omega_{g,q}^2}-c^2_s)\frac{\omega_{g,q}}{T}(\nabla\cdot\vec{u})\bigg],\nonumber\\ 
\end{eqnarray}
where $c^2_s$ is the speed of sound, $m^2_D$ is the Debye mass, ${\varepsilon}$ is the energy density,  $\tau_{el}$ is the time scale
of the instability in chromo-electric fields.  The expression is mathematically similar to~\cite{bmuller}. The only difference is the appearance of modified 
quasi-particle dispersion.

 Finally, the effective Vlasov-Dupree equation (linearized) in the presence of 
turbulent color fields with the above ansatz is formulated in Refs.~\cite{bmuller, chandra_eta} reads: 
\begin{eqnarray}
\label{eqt}
 &&\bigg\lbrace (\frac{p^2}{3 \omega_{g,q}}-c_s^2)\frac{\omega_{g,q}}{T} (\nabla\cdot \vec{u})+\frac{p_i p_j (\nabla u)_{ij}}{\omega_{g,q} T}\bigg\rbrace f_0^{g,q} (1\pm f_0^{g,q})=\nonumber\\
 &&\frac{g^2 C_2}{3 (N_c^2-1) \omega^2_{g,q}} \langle E^2+B^2 \rangle \tau_m {\mathcal L}^2 f_1^{g,q} (\vec{p}) f_0^{g,q} (1\pm f_0^{g,q}).
 \end{eqnarray}
Importantly, first term in the left hand side of Eq.(\ref{eqt}) contributes to the physics of
isotropic expansion (bulk viscosity effects) which is not taken into account in the present work. As the analysis is valid
for temperatures which are away from $T_c$, the bulk viscous effects can conveniently be neglected there. 

Next, solving  Eq.~(\ref{eqt}) for $\Delta$ analytically, we obtain the following expression~\cite{chandra_zeta},

\begin{eqnarray}
\label{eqdel}
\Delta (\vec{p})&=&2 (N_c^2-1)\frac{\omega_{g,q} T}{3 C_2 g^2 \langle E^2+B^2 \rangle_{g,q} \tau_m}.
\end{eqnarray}

 The unknown factor, $\langle E^2+B^2 \rangle_{g,q} \tau_m$ in the denominator  of Eq. (\ref{eqdel}) can be related to the 
phenomenologically known quantity the jet quenching parameter, $\hat{q}$, in both gluonic and quark sectors as done in Ref.~\cite{majumder}.  
This connection is established as mentioned below. The two crucial transport coefficients that may get significant contribution from the 
turbulent color fields are shear viscosity, $\eta$ and jet quenching parameter, $\hat{q}$. Thus, the 
strength of momentum anisotropy in the expanding QGP medium  can be related to the physics of these parameters.
Recall that the strength of the anisotropy, $\Delta(\vec{p})$ is  related to the  $\eta$. The coefficient,  $\eta$ is seen to be 
inversely proportional to the $\hat{q}$~\cite{bmuller, chandra_eta}. The  jet quenching parameter, $\hat{q}$ turns out to be proportional to the mean momentum square per unit length 
on the an energetic parton imparted by turbulent fields~\cite{ask:2010}. In that context the anisotropy is related to the 
quenching. Here,  we relate the unknown quantities $\langle E^2+B^2 \rangle \tau_m$- which encodes the physics of 
anisotropy and chromo-Weibel instability,
to the $\hat{q}$ both in gluonic and matter sectors as \cite{majumder}
\begin{equation}
\label{eqhat}
 \hat{q}= \frac{2 g^2 C_{g/f}}{3(N_c^2-1)} \langle E^2+B^2 \rangle \tau_m,
\end{equation}
where $C_g= N_c$, $C_f=\frac{(N_c^2-1)}{2 N_c}$ for the gluons and quarks respectively. In the present case, we have $N_c=3$.

Employing the definition of $\hat{q}$ from Eq. (\ref{eqhat}) in Eq. (\ref{eqdel}), we obtain the following expression for 
the $\Delta$ term,
\begin{eqnarray}
 \Delta &=& \frac{4 \omega_{g,q}^2 T}{9 \hat{q}_{g,q}}\nonumber\\
\end{eqnarray}

With the above relation, we obtain the near equilibrium distribution functions in terms of the jet quenching parameter $\hat{q}$ as,
\begin{eqnarray}
\label{eq20}
 f^{g,q} (\vec{p})&=&f_0^{g,q}-f_0^{g,q} (1 \pm f_0^{g,q}) \frac{4 \omega_{g,q}}{9 \hat{q}_{g,q} T\tau} \big(p_z^2-\frac{p^2}{3}\big).
 \end{eqnarray}

The jet quenching parameter,  $\hat{q}$ for both gluons and quarks has been estimated phenomenologically within several different approaches~\cite{burke,55,50,51,56,42}.
The gluon quenching parameter $\hat{q}_g$ 
can be obtained from the relation, $\hat{q}_g \equiv \frac{9}{4}$ (in terms of Casimir invariants of the $SU(3)$ group). As we shall see later, such values of
$\hat{q}_{g,q}$ induce very strong non-perturbative effects to the dileptons spectra.

It is to be noted that we have only three independent functions ($z_g$, $z_q$, and  $\hat{q}$) ($\hat{q}$ for gluons and quarks are related by the respective  Casimirs of $SU(3)$). The first two functions are 
estimated while describing the hot QCD equation of state. As stated earlier,  the third one is a phenomenological parameter. Let us now proceed to investigate the 
thermal particle rates in the QGP medium.

It is important to note that momentum anisotropy and plasma instabilities in the initial stages of the RHIC have  extensively been studied in the literature by 
several authors while investigating the effects of color collective excitations~\cite{smrow} and thermalization~\cite{arnold1}  there.  Rebhan {\it et. al},~ \cite{rebhan} studied the 
dynamics of non-abelian plasma instabilities within Hard-loop dynamics and Romatschke and Venugopalan~\cite{venu} explored  the same within CGC framework. The collective plasma modes in anisotropic QGP  have been 
studied in~\cite{roma1,carrig,chandra_ins}. The collisional energy loss in anisotropic QGP~\cite{roma2}, the momentum broadening~\cite{roma3}, the radiative energy loss~\cite{roy}, and  the wake potential~\cite{wake} 
are some of the related effects those have been investigated in the anisotropic QGP.  For recent  reviews, we refer the reader to  Refs.~\cite{strik_pr,rev1} and references therein.

\section{Modified Thermal particle rates in QGP}
Particles are produced in all stages during  fireball evolution in the heavy-ion collisions from locally thermalized QGP medium.
The main focus here is to investigate dilepton pair in the thermal QGP medium via $q\bar{q}$ -annihilation (most dominant source of 
dilepton pair production). As discussed in detail earlier, 
the momentum anisotropy present during the hydrodynamic evolution of the QGP medium in RHIC can be 
captured as the modification in the equilibrium (local) 
distributions of the gluonic and quark-antiquark degrees of freedom by setting up and solving 
an appropriate Vlasov-Boltzmann equation (linearized). The linear approximation is valid while the modifications 
are much smaller as compared to their equilibrium counterpart. In this near equilibrium situation, we can still utilize the 
kinetic theory for obtaining particle production rates and yields. This could be done by just replacing the expressions for the momentum distributions as the modified ones as done below.

The major source of thermal dileptons in the QGP medium is the $q\bar q$ annihilation process, 
$q\bar q\rightarrow \gamma^* \rightarrow l^+ l^-$.  Rate of dilepton production for this process can be written as \cite{rvogt}:
\ba
 \frac{dR}{d^4p}^{\tiny{l^+l^-}}&=&\int \left[g f(\vec{p}_1) \frac{d^3\vec{p}_1}{(2\pi)^3}\right] 
 \left[g f(\vec{p}_2)\frac{d^3\vec{p}_2}{(2\pi)^3}\right] \nonumber \\
 && \times \sigma(M^2)\, v_{rel}\, \delta^4(p-p_1-p_2). \label{dilrateEq}
\ea
Here $p=(E=E_1+E_2,\,\vec{p}=\vec{p}_1+\vec{p}_2)$ is the four momentum of the dileptons 
and $p_{1(2)}=(E_{1(2)},\vec{p}_{1(2)})$ is that of quark or anti-quark with  
$E_{1(2)}\simeq|\vec{p}_{1(2)}|$, while neglecting the quark masses. 
We denote the invariant mass of the virtual photon as 
$M^2=(E_1+E_2)^2-(\vec{p}_1+\vec{p}_2)^2$. 
 The relative velocity of the quark-anti-quark pair is given by 
 $v_{rel}=\sqrt{\frac{M^2(M^2-4m_q^2)}{4E_1^2E_2^2}}\simeq \frac{M^2}{2E_1E_2}$ 
 and 
the term $\sigma(M^2)$ is the relevant thermal dilepton production cross section. 
With $N_f$=2 and $N_c=3$, we have $M^2g^2\sigma(M^2)=\frac{80\pi}{9}\alpha^2$ \cite{Alam:1996fd}. 
The functions $f(\vec{p})$ are the quark and anti-quark distribution functions with $g$ being 
the corresponding degeneracy factor. 
The expressions for quark and antiquark distribution functions are obtained in the previous 
section, have the expressions: 
\ba
f (\vec{p})=f_0-f_0 
(1 - f_0) \frac{4 \omega_{q}}{9 \hat{q}_{q} T} \big(p_i p_j \Delta u_{ij}\big).
\ea
Since, we are interested in large invariant mass regime $M\gg T\gg m_q$, we can approximate,
Fermi-Dirac distribution by classical Maxwell-Boltzmann distribution, 
so that
\ba\label{MxBzn}
  f_0(\vec{p_i}) &\approx& z_q e^{-E_i/T} \nn
  f(\vec{p_i}) &\approx& z_q e^{-E_i/T} \left[ 1 - \frac{4 \omega_{q}}{9 \hat{q}_{q} T} \big(p_i p_j \Delta u_{ij}\big)\right].
\ea

Keeping only  up to the quadratic order in momenta, we can write the dilepton 
production rate as\cite{Bhatt:2011kx}, 

\ba
 \frac{dR}{d^4p}^{\tiny{l^+l^-}}&=&\int \frac{d^3\vec{p}_1}{(2\pi)^3} \frac{d^3\vec{p}_2}{(2\pi)^3} 
 \frac{M^2g^2\sigma(M^2)}{2E_1E_2}\delta^4(p-p_1-p_2) \nonumber \\
 && \times f_0(\vec{p_1})f_0(\vec{p_2}) \left[1-2\, \frac{4 \omega_{q}}{9 \hat{q}_{q} T} 
 \big({p_1}_i {p_1}_j \Delta u_{ij}\big) \right]  \nonumber \\
 &=& \frac{dR_0}{d^4p}^{\tiny{l^+l^-}} +\, \frac{dR_1}{d^4p}^{\tiny{l^+l^-}}. \label{R}
\ea
The equilibrium contribution to the dilepton production 
\ba 
\frac{dR_0}{d^4p}^{\tiny{l^+l^-}}&=&\int \frac{d^3\vec{p}_1}{(2\pi)^6}~\frac{M^2g^2\sigma(M^2)}{2E_1E_2} \delta(E-E_1-E_2)
\nonumber \\
&& \times ~z_q^2 e^{-(E_1+E_2)/T}
\ea
is known and is given by\cite{Chandra:2015rdz} 
\ba\label{R0}
\frac{dR_0}{d^4p}^{\tiny{l^+l^-}}&=&\frac{z_q^2}{2}~\frac{M^2g^2\sigma(M^2)}{(2\pi)^5}~e^{-E/T}.
\ea
Next, we calculate the non-equilibrium contribution to dilepton production, given as 

\ba
\frac{dR_1}{d^4p}^{\small{l^+l^-}}&=&-2\int \frac{d^3\vec{p}_1}{(2\pi)^6} ~\frac{M^2g^2\sigma(M^2)}{2E_1E_2} \delta(E-E_1-E_2)\nonumber \\
&& ~~~~\times ~z_q^2 e^{-(E_1+E_2)/T}\left[ \frac{4 \omega_{q}}{9 \hat{q}_{q} T} 
 \big({p_1}_i {p_1}_j \Delta u_{ij}\big) \right]\nonumber \\
 &=& I_{ij}(p) \Delta u_{ij}.
\ea
Here we have represented 
\ba \label{I}
I_{ij}(p)&=&-2\int \frac{d^3\vec{p}_1}{(2\pi)^6} ~\frac{M^2g^2\sigma(M^2)}{2E_1E_2} \delta(E-E_1-E_2)\nonumber \\
&& ~~~~\times ~z_q^2 e^{-(E_1+E_2)/T} \frac{4 \omega_{q}}{9 \hat{q}_{q} T} 
 {p_1}_i {p_1}_j . 
\ea
Now, the most general form of this second rank tensor in LRF can be written as, 
\be
I_{ij}(p)=a_0~\delta_{ij} + a_2~ p_ip_j.
\ee
However, we note that on contraction with $\Delta u_{ij}$ only $a_2$ will survive as $\delta_{ij}\Delta u_{ij}=0$. 
By constructing projection operator $Q_{ij}$ we can calculate $a_2=Q_{ij}I_{ij}$, where form of the projection 
operator in LRF is given by
\be \label{pro-op}
Q_{ij}=-\frac{1}{2|{\vec{p}}|^2}\left(\delta_{\ij}-\frac{3p_ip_j}{|{\vec{p}}|^2}\right).
\ee
Now, the expression for non-equilibrium contribution to dilepton rate becomes, 
\ba
\frac{dR_1}{d^4p}^{\small{l^+l^-}}&=&I_{ij}(p) \Delta u_{ij}
=\left[Q_{mn}I_{mn}\right] p_i p_j \Delta u_{ij}. \label{R1d}
\ea
Computations of $Q_{mn}I_{mn}$ can be done using Eq.(\ref{I}) and Eq.(\ref{pro-op}), 
\ba
Q_{mn}I_{mn}&=&-2\int \frac{d^3\vec{p}_1}{(2\pi)^6} ~\frac{M^2g^2\sigma(M^2)}{2E_1E_2} \delta(E-E_1-E_2)\nonumber \\
&& ~~~~\times ~z_q^2 e^{-(E_1+E_2)/T} \frac{4 \omega_{q}}{9 \hat{q}_{q} T} 
 {p_1}_m {p_1}_n \nn
 && ~~~~ \times \left[ -\frac{1}{2|{\vec{p}}|^2}\left(\delta_{mn}-\frac{3p_mp_n}{|{\vec{p}}|^2}\right)\right] \nonumber \\
 &=& -\frac{8}{9}\frac{1}{\hat{q}_{q} T}\int \frac{d^3\vec{p}_1}{(2\pi)^6} ~\frac{M^2g^2\sigma(M^2)}{2E_1E_2} \delta(E-E_1-E_2)\nonumber \\
&& ~~~~~~~~~~\times ~z_q^2 e^{-(E_1+E_2)/T} \{p_1+T^2\partial_T\, ln(z_q)\} \nn
 && ~~~~~~~~~~ \times \left[ -\frac{|{\vec{p_1}}|^2}{2|{\vec{p}}|^2}+\frac{3(\vec{p}\cdot \vec{p_1})}{2|{\vec{p}}|^4}\right] \nn
 &=& -\frac{8}{9}\frac{1}{\hat{q}_{q} T} \left[\mathscr{M}+\mathscr{N} \right], \label{QI}
 \ea
where we have used Eq.(\ref{eq2}): $\omega_q=p+T^2\partial_T\, ln(z_q)$. Next, we calculate integrals $\mathscr{M}$ and $\mathscr{N}$, 
and they are found to be 
\ba
\mathscr{M}&=& \int \frac{d^3\vec{p}_1}{(2\pi)^6} ~\frac{M^2g^2\sigma(M^2)}{2E_1E_2} \delta(E-E_1-E_2)\nonumber \\
&& \times ~z_q^2 e^{-(E_1+E_2)/T} \{p_1\} \left[ -\frac{|{\vec{p_1}}|^2}{2|{\vec{p}}|^2}
+\frac{3(\vec{p}\cdot \vec{p_1})}{2|{\vec{p}}|^4}\right] \nn
&=& \frac{E}{4}~\frac{dR_0}{d^4p}^{\tiny{l^+l^-}}
\ea
and
\ba
\mathscr{N}&=& \int \frac{d^3\vec{p}_1}{(2\pi)^6} ~\frac{M^2g^2\sigma(M^2)}{2E_1E_2} \delta(E-E_1-E_2)\nonumber \\
&& \times ~z_q^2 e^{-(E_1+E_2)/T} \{T^2\partial_T\, ln(z_q)\} \left[ -\frac{|{\vec{p_1}}|^2}{2|{\vec{p}}|^2}
+\frac{3(\vec{p}\cdot \vec{p_1})}{2|{\vec{p}}|^4}\right] \nn
&=& \frac{1}{3}~T^2\partial_T\, ln(z_q)~\frac{dR_0}{d^4p}^{\tiny{l^+l^-}}.
\ea
Non-equilibrium contribution to dilepton rates can be written using Eq.(\ref{QI}) and Eq.(\ref{R1d}) as
\ba
\frac{dR_1}{d^4p}^{\small{l^+l^-}} &=& -\frac{8}{9}\frac{1}{\hat{q}_{q} T}\left[\frac{E}{4}
+\frac{1}{3}T^2\partial_T\, ln(z_q)\right]  \nn
&& \times \frac{dR_0}{d^4p}^{\tiny{l^+l^-}}p_i p_j \Delta u_{ij}. 
\ea
Finally, we write the expression for \textit{total} dilepton production, using Eq.(\ref{R}) as 
\ba
\frac{dR}{d^4p}^{\small{l^+l^-}}&=&\left[1 -\frac{8}{9}\frac{1}{\hat{q}_{q} T}\left(\frac{E}{4}
+\frac{1}{3}~T^2\partial_T\, ln(z_q)\right)p_i p_j \Delta u_{ij}\right]\nn
&& \times \frac{dR_0}{d^4p}^{\tiny{l^+l^-}}.
\ea

\begin{figure}
\includegraphics[scale=0.6]{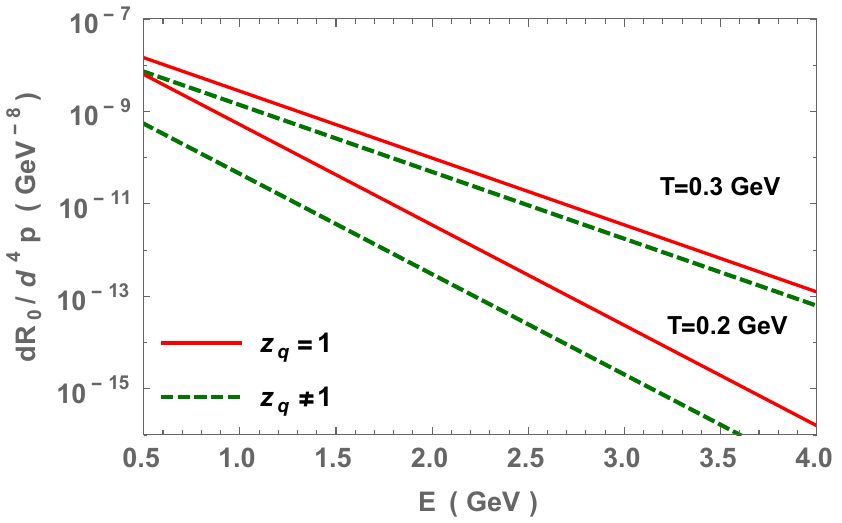}
\caption{
The equilibrium dilepton production rates with 
of $z_q^2$ as function of $T/T_c$ is shown along with its SB limit ($z_q\rightarrow 1$). The temperature dependence of the 
effective quark fugacity, $z_q$ is taken from Ref.~\cite{chandra_quasi}.}
\label{fig1}
\end{figure}

Next, we plot the equilibrium rates $\frac{dR_0}{d^4p}^{\small{l^+l^-}}$ using Eq.(\ref{R0})
for two different temperatures (T=0.3 GeV and T=0.2 GeV) in Fig. \ref{fig1} with 
effective quark fugacity $z_q$ taken from Ref.~\cite{chandra_quasi} (dotted line) and with $z_q=1$ (solid line). 
The latter case corresponds to equation of state of ultra-relativistic massless quarks and gluons (\textit{ideal}). From the 
above figure it is clear that the effect of $z_q$ is to suppress the rates uniformly for at all dilepton energies 
and suppression is more dominant at lower temperatures \cite{Chandra:2015rdz}. 

Authors had calculated the effect of realistic equation of state, via $z_q$ 
on dilepton production (Eq.(\ref{R0})), along with the effect of shear and bulk viscosities 
in Ref. \cite{Chandra:2015rdz}. 
Dilepton production rate expression obtained 
in this work brings out the effect of turbulence and momentum anisotropy, 
apart from the equation of state. 
Though the method of getting the rates remains same, 
the non-equilibrium effect included in the distribution functions used 
in Ref. \cite{Chandra:2015rdz} is that of viscosities, unlike 
that of turbulent chromo fields in the present work. 

It is be noted that all above analysis were done in the rest frame of the medium, therefore 
in general frame with four-velocity $u^\mu$, these results become, 
\ba\label{finalR}
\frac{dR_0}{d^4p}^{\small{l^+l^-}}&=&\frac{z_q^2}{2}~\frac{M^2g^2\sigma(M^2)}{(2\pi)^5}~e^{-u\cdot p/T},\\
\frac{dR_1}{d^4p}^{\small{l^+l^-}}&=&-\frac{8}{9}\frac{1}{T^4}\left( \frac{\hat{q}_{q}}{T^3} \right)^{-1}\left[\frac{u\cdot p}{4}
+\frac{1}{3}~T^2\partial_T\, ln(z_q)\right]\nn
&& \times \frac{dR_0}{d^4p}^{\tiny{l^+l^-}}\left[p^\mu p^\nu \Delta u_{\mu\nu}\right],\,\nonumber
\ea
where we have kept the term $\hat{q}_q/T^3$ for phenomenological reasons stated in the previous session. 

\par
We now proceed to study the QGP thermal dilepton spectra from heavy-ion collisions with non-equilibrium contributions.

\section{Thermal dilepton yield from QGP during fireball evolution}
To study dilepton yield from the QGP phase in heavy-ion collisions, we need to model 
expansion of the thermalised fireball. This can be done using relativistic hydrodynamics. 
In this qualitative analysis, we use the longitudinal boost invariant flow model of Bjorken\cite{Bjorken:1982qr} 
to describe the expanding system. In the \textit{Bjorken flow}, with 
the parametrization $t=\tau$ cosh$\,\eta_s$ and $z=\tau$ sinh$\,\eta_s$; 
with the proper time $\tau = \sqrt{t^2-z^2}$ and space-time 
rapidity $\eta_s=\frac{1}{2}\,ln[\frac{t+z}{t-z}]$; the four velocity of the medium is written as 
$u^\mu=(\rm{cosh}\,\eta_s,0,0,\rm{sinh}\,\eta_s)$. Neglecting the 
effects of viscosity, now we can write the energy dissipation equation for the system as\cite{Bjorken:1982qr}
\begin{equation}\label{hydro}
 \frac{d\varepsilon}{d\tau}+\frac{\varepsilon + P}{\tau}=0.
\end{equation}
Here $\varepsilon$ is the energy density and $P$ is the pressure of the system. Above equation 
need to be closed by providing equation of state (EoS). We use recent lattice QCD EoS \cite{cheng} for this purpose. 
We take the transition temperature $T_C$, denoting the end of QGP phase, as $180$ MeV in this analysis. 
By providing the initial conditions i.e.; $\tau_0=0.5$ fm/c and $T(\tau_0)=300$ MeV relevant for RHIC energies, we now solve the 
energy dissipation equation numerically to obtain the temperature profile $T(\tau)$.

Equipped with the temperature dependent thermal dilepton production rates, \textit{dilepton yield} from the 
QGP can  be obtained by integrating these rates over the space-time history of the fireball evolution, 
\be
\frac{dN}{d^4p}^{\small{l^+l^-}}=\int d^4x \, \frac{dR}{d^4p}^{\small{l^+l^-}}.
\ee
The four-volume element within Bjorken model is given by $d^4x=\pi R_A^2d\eta_s\tau d\tau$. 
Here $R_A=1.2 A^{1/3}$ is the radius of the nucleus used for the collision and for $Au,~ A=197$. 
We parametrise the four momentum of the dilepton as 
$p^{\alpha}$ = $(m_T coshy,p_T cos\phi_p,p_T sin\phi_p,m_Tsinhy)$ with $m_T^2$ = $p_T^2+M^2$. 
The factors appearing in the rate expressions i.e.; Eq. (\ref{finalR}) to be used in above integral are given as 
$u.p=m_T cosh(y-\eta_s)$ and  
\begin{eqnarray}\label{vf}
p^{\alpha}p^{\beta} \Delta u_{\mu\nu} &=& -\frac{\,M^2}{3\tau}-\frac{m_T^2}{\tau}\,sinh^2(y-\eta_s).
\end{eqnarray}

Desired dilepton yields in terms of invariant mass $M$, transverse momentum $p_T$ and momentum rapidity $y$ 
are now given by 
\begin{eqnarray}\label{tot-yield}
\frac{dN^{\tiny{l^+l^-}}}{dM^2d^2p_Tdy}
&=&\pi R_A^2\int_{\tau_0}^{\tau_f}
d\tau ~\tau \int_{-\infty}^\infty d\eta_s\,
\frac{1}{2}\frac{dR}{d^4p}^{\tiny{l^+l^-}}\nonumber \\
&=&\frac{dN_0^{\tiny{l^+l^-}}}{dM^2d^2p_Tdy} + \frac{dN_1^{\tiny{l^+l^-}}}{dM^2d^2p_Tdy} .
\end{eqnarray}
After performing the $\eta_s$ integration, the equilibrium and non-equilibrium contributions to the 
total dilepton yield are obtained as,
\ba\label{NpT}
\frac{dN_0^{\tiny{l^+l^-}}}{dM^2d^2p_Tdy}&=&\mathscr{R}
\int_{\tau_0}^{\tau_f} \tau\,d\tau\, z_q^2\,2K_0(z_m), \\
\frac{dN_1^{\tiny{l^+l^-}}}{dM^2d^2p_Tdy}&=&-\frac{8}{9}\mathscr{R}\,
\int_{\tau_0}^{\tau_f} d\tau\, \frac{z_q^2}{T^4}\, \left( \frac{\hat{q}_{q}}{T^3} \right)^{-1} 
\mathscr{T}(T, p),\nn
{\rm with},\, 
\mathscr{T}(T, p)&=& K_0(z_m)\,\bigg\lbrace \frac{1}{3}T^2\partial_T\,ln(z_q)\left(m_T^2-\frac{2}{3}M^2\right) \bigg\rbrace \nn
&&+K_1(z_m)\,\bigg\lbrace\frac{m_T^3}{8}-\frac{1}{6}M^2m_T \bigg\rbrace \nn
&&-K_2(z_m)\,\bigg\lbrace\frac{1}{3}T^2\partial_T\,ln(z_q)\,m_T^2\bigg\rbrace \nn
&&-K_3(z_m)\,\bigg\lbrace\frac{m_T^3}{8}\bigg\rbrace,\nn
{\rm and},\,\,\,\mathscr{R}&=&\left[\frac{\pi R_A^2}{2^2}\frac{1}{(2\pi)^5}\frac{80\pi\alpha^2}{9}\right].\nonumber 
\ea
 
Here 
$K_n$ are the modified Bessel functions of the second kind and $z_m\equiv m_T/T$. 
Now we numerically integrate the above integrals with temperature profile obtained from 
hydrodynamical analysis to get the dilepton yields. All the results are presented for the 
midrapidity region of the dilepton i.e.; $y=0$. 

\begin{figure}
\includegraphics[scale=0.42]{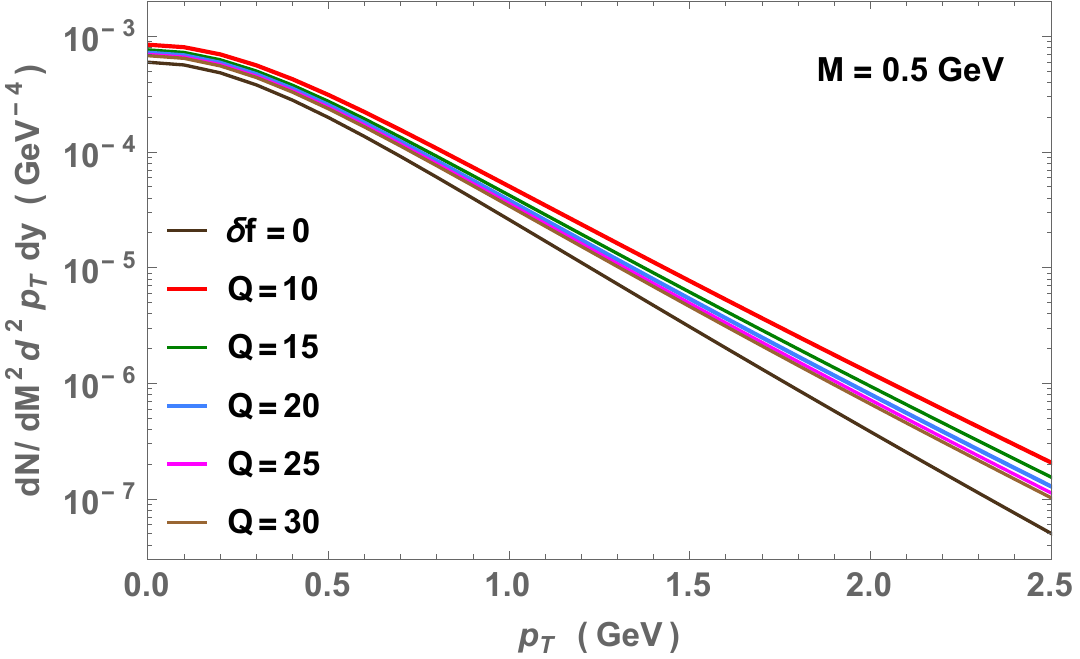}
\caption{
Dilepton yields with non-equilibrium effects for different $\hat{q}_q/T^3\equiv Q$ values for $M=0.5$ GeV. 
Equilibrium yield contribution is also plotted ($\delta f=0$).}
\label{fig2}
\end{figure}

The thermal dilepton yields as a function of transverse momentum of the dileptons for the invariant mass $M=0.5$ GeV are shown in 
Fig. \ref{fig2}. The non-equilibrium effects are included with various jet-quenching parameter $\hat{q}_q/T^3$ values. The equilibrium 
contribution alone is also plotted ($\delta f=0$) for comparison. It can be seen that the effect of 
non-equilibrium terms is to enhance the dilepton spectra throughout the $p_T$ regime. 
Also, as we increase the $\hat{q}_q/T^3$ value, yield decreases and approaches the equilibrium value. 
From Eqs.(\ref{finalR}) \& (\ref{vf}), it is clear that non-equilibrium contribution to dilepton 
rates is addictive, hence we see an increase in the yields with the inclusion of non-equilibrium terms. 

Notably, with $\hat{q}_q/T^3=15$, we observe $\sim38\%$ enhancement at $p_T=0.5$ GeV and $\sim147\%$ at $p_T=2$ GeV. 
Since, $\hat{q}_q/T^3$ term appear in denominator of the yield expression, as seen in Eq. (\ref{NpT}), 
increasing its value will result in the decrease of non-equilibrium contribution. 
For e.g.; for $\hat{q}_q/T^3=25$, enhancement is only about $\sim23\%$ and $\sim88\%$ for transverse momenta
0.5 GeV and 2 GeV respectively.

It is crucial to note that enhancement of the spectra is more significant at high $p_T$, indicating 
the strong non-equilibrium effects at that regime. 
In heavy-ion collisions, high $p_T$ particles are produced during the initial stages of the evolution. 
Since we expect the anisotropic effects also to be dominant at the initial stages of the evolution, 
its effect will be more effective at high $p_T$. The fact that enhancement is seen in low $p_T$ particles 
indicate that the non-equilibrium effects remain significant throughout the evolution of the system. 
We note that just like the overall effect of viscosities as seen in Ref.\cite{Chandra:2015rdz}, 
present non-equilibrium effect also enhances the thermal dilepton spectra. 
\begin{figure}
\includegraphics[scale=0.42]{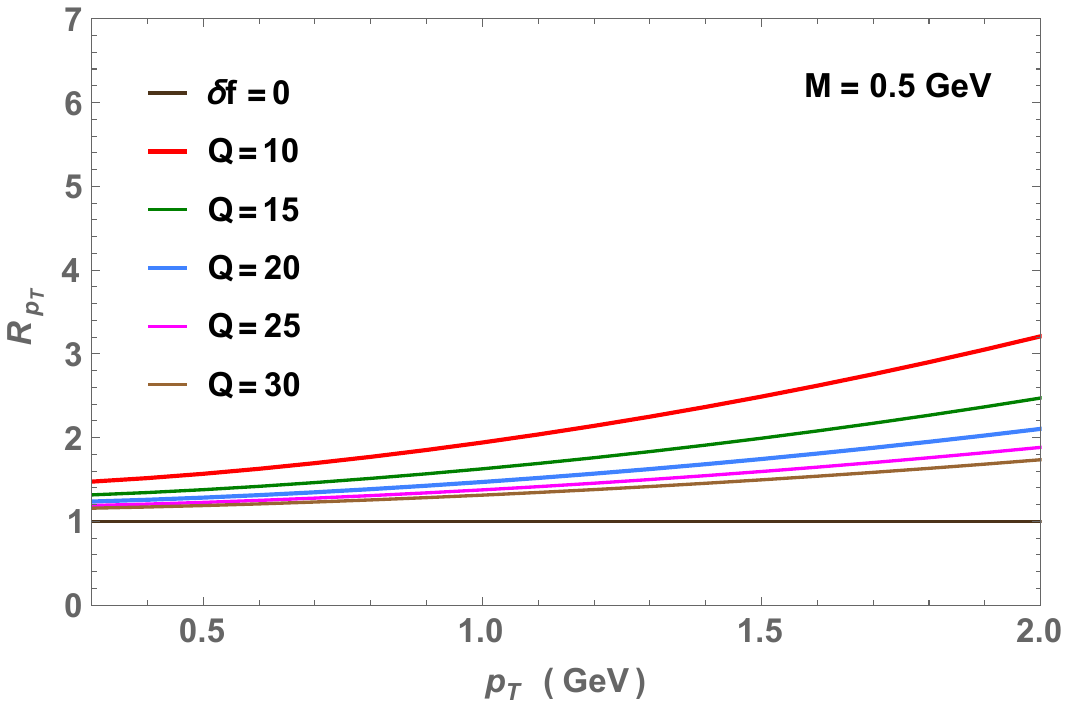}
\caption{
Strength of non-equilibrium corrections to equilibrium yield represented through ratio $R_{p_T}$ (Eq. (\ref{Rpt})) 
for different jet quenching parameters.} 
\label{fig3}
\end{figure}

Next, we study the strength of these non-equilibrium corrections to equilibrium distribution functions by 
looking into their contributions to dilepton spectra. We begin our analysis 
by constructing the following ratio 
\begin{equation}\label{Rpt}
R_{p_T}=\frac{dN^{\tiny{l^+l^-}}}{dM^2d^2p_Tdy}/\frac{dN_0^{\tiny{l^+l^-}}}{dM^2d^2p_Tdy},
\end{equation}
where numerator includes non-equilibrium contributions. We plot this ratio as a function of 
transverse momenta of the dileptons for different $\hat{q}_q/T^3$ values in Fig. \ref{fig3}. 

Note that, for smaller value of transverse momentum, the non-equilibrium contribution 
is changing the equilibrium part  by $\sim50\%$ for $\hat{q}_q/T^3>10$. 
These non-equilibrium contributions tend to increase strongly as we move towards higher, $p_T$. 
The corrections start decreasing as we increase the $\hat{q}_q/T^3$ parameter 
as expected, since high values of $\hat{q}_q/T^3$ dilutes non-equilibrium corrections.
For $\hat{q}_q/T^3=20(30)$, the contribution begins with $\sim 28(19)\%$ at $p_T=0.5$ GeV and reaches to $\sim110(74)\%$ 
by $p_T=2$ GeV. Overall, we observe strong corrections to the yield by non-equilibrium effects.

It is to be emphasised that for low values of $\hat{q}_q/T^3$, 
significant corrections to 
spectra are seen. These strong corrections due to perturbative non-equilibrium effects 
are indicative of the fact that such $\hat{q}_q/T^3$ values are preferably ruled out within the present model. 
However to substantiate the claim thoroughly, one may need to perform a quantitative analysis including 
three-dimensional hydrodynamical flow, which is beyond the scope of this paper. 
Moreover, we recall that, in the present work, the corrections to spectra are calculated within one dimensional 
Bjorken flow, which is known to overestimate the particle spectra. So the corrections 
shown in this qualitative study act only as upper bounds. The precise nature of corrections to particle yields 
depends very much upon the geometry under consideration (here, simplified Bjorken), because of the involved 
space-time integration (which we perform numerically). How a different, more realistic three-dimensional geometry 
involving transverse flow will change the overall corrections, cannot be guessed. 
Although, it is expected that the effect of transverse flow is to decrease the 
particle yield, since the evolution time and therefor the limit of proper time integration, 
will be less in that case. 
However, the space part integration contribution 
is non-trivial and may change significantly under a different geometry. 
 \begin{figure}
 \includegraphics[scale=0.42]{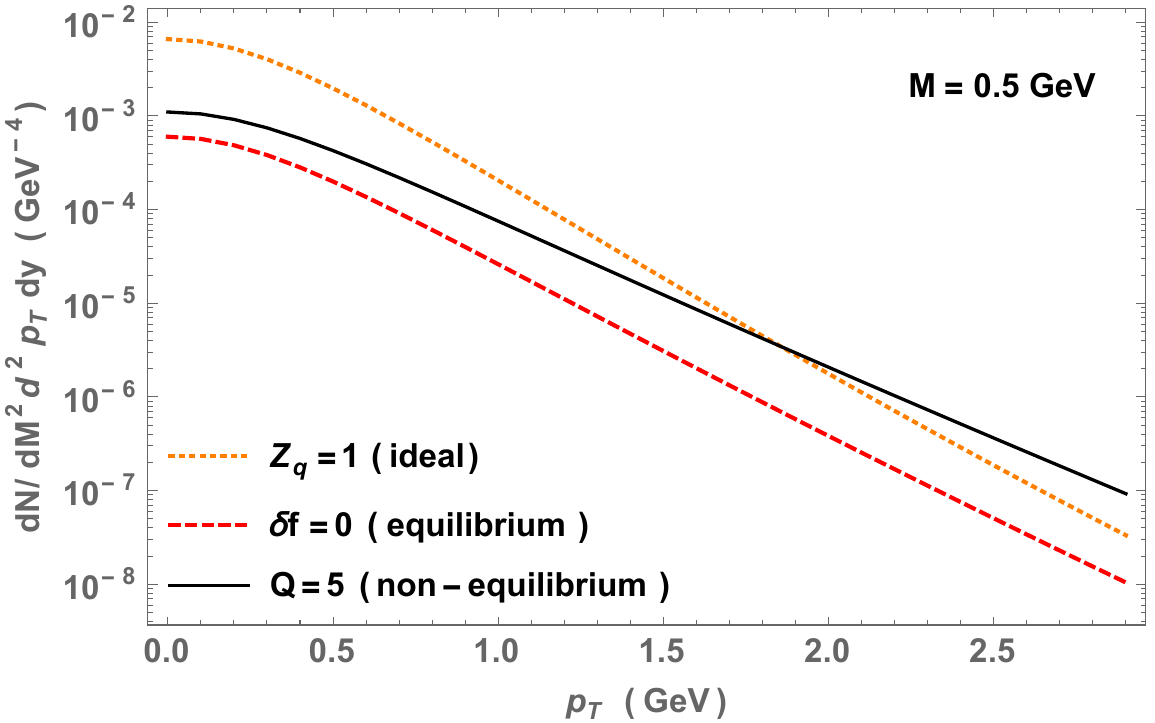}
 \caption{
 Strength of non-equilibrium terms via ratio $R_{p_T}$ }
 \label{fig4}
 \end{figure}

Finally, let us analyze, how the non-equilibrium corrections and equilibrium values are modifying 
the \textit{ideal} case i.e.; $z_q=1$. 
Yield corresponding to $ideal$ case can be obtained by considering Eq.(\ref{R0}) with $z_q=1$. 
The effect of $z_q^2$ term on the ideal spectra 
was analyzed by the authors in detail elsewhere \cite{Chandra:2015rdz} and wont be repeated here. 
We recall that it's effect was found to be suppressive and it can also be inferred from Fig. \ref{fig1} 
of this manuscript. In Fig. \ref{fig4}, we plot the equilibrium and non-equilibrium dilepton yield for 
$M=0.5$ GeV with that for the \textit{ideal} case. For non-equilibrium part, we have taken the 
jet quenching parameter value as $\hat{q}_q/T^3=5$ (This is typically the value that we obtain by averaging the predictions of all these approaches~\cite{burke,55,50,51,56,42}). For low transverse momenta, non-equilibrium enhancement 
of equilibrium spectra is marginal compared to the $ideal$ values. 
We observe that the non-equilibrium effects can overtake the $z_q$ suppressions of ideal spectra ($z_q=1$) 
at high $p_T$s. At at $p_T\sim2$ GeV, we have near cancellation of these effects. However, thereafter 
non-equilibrium contribution will dominate other two. However, 
by increasing the jet quenching parameter, strength of the $z_q$ suppressions can be made dominate so that 
total spectra will become less than less than $ideal$ case. 

It is to be emphasized that  in the present, qualitative study, we have used 
one-dimensional Bjorken flow to model the system. 
It is  well known that Bjorken results tend to over-estimate the particle production because 
system takes more  time to cool down compared to realistic three-dimensional expansion 
with the transverse flow~\cite{Bhatt:2011kx}. However, such a quantitative study is not within the scope of present analysis and will be 
taken up for investigation the near future. It is encouraging that we are seeing very significant effects with one-dimensional flow, which, we believe,  
may  guide us to have observable signatures on realistic three-dimensional calculations.

\section{Conclusions and outlook}
In conclusion, thermal particle production is investigated in the presence of momentum anisotropy during the hydrodynamical expansion of the QGP in 
heavy-ion collisions. The effects of anisotropy are encoded in the non-equilibrium part of the quark/anti-quarks and gluon momentum distribution functions.
We particularly, studied the dilepton production rate and compared our results against the isotropic/equilibrium case. The modifications 
induced by the anisotropy are found to be significant as far as the rate and dilepton yields are concerned. The strength of anisotropy,
which in our case, is inversely proportional to the jet quenching parameter apart from other momentum dependent factors, have appreciable
impact on the rate and yield. The whole analysis is based on  an effective transport equation which is obtained by an ensemble averaging of the 
turbulent gluonic fields created due to the momentum anisotropy by inducing instability in $SU(3)$ Yang-Mills equations.
We can perhaps treat  the thermal particle production 
in the presence of the  momentum  anisotropy as the indicator of the impact of turbulent color-fields and anomalous transport processes
during the expansion of QGP.

It would be interesting to include collisional processes and study the interplay of the them with anomalous ones by setting up and solving appropriate 
transport equation for the non-equilibrium distributions.   We intend to employ them to study thermal particle production in heavy-ion collisions.

\section*{Acknowledgements} 
This work has been conducted under the INSPIRE Faculty 
grant of Dr. Vinod Chandra ({\tt Grant no.:IFA-13/PH-55, Department of Science and Technology, Govt. of India}) at Indian Institute of Technology Gandhinagar, India. 
We would like to record our sincere gratitude to the people of India for their generous support for the research in basic sciences in the country.

\end{document}